\documentclass[pre,aps,reprint, superscriptaddress]{revtex4-1}
\usepackage{amsmath}
\usepackage{graphicx}
\usepackage{subfigure} 
\usepackage{fancyhdr}
\begin{document}
\title{Pendular Behavior of Public Transport Networks}
\author{Mirian M. Izawa}
\affiliation{Instituto de Física - Universidade de Bras\'ilia, 70919-970,  Brazil}
\author{Fernando A. Oliveira}
\affiliation{Instituto de Física - Universidade de Bras\'ilia, 70919-970,  Brazil}
\author{Daniel O. Cajueiro}
\affiliation{Departamento de Economia - Universidade de Bras\'ilia, 70919-970, Brazil}
\author{Bernardo A. Mello}
\affiliation{Instituto de Física - Universidade de Bras\'ilia, 70919-970,  Brazil}
\affiliation{IBM Thomas J. Watson Research Center, Yorktown Heights, New York 10598, USA}

\date{June 1, 2017 }

\begin{abstract}
In this paper, we propose a methodology that bears close resemblance to the Fourier analysis of the first harmonic to study networks subjected to pendular behavior.  In this context, pendular behavior is characterized by the phenomenon of people's dislocation from their homes to work in the morning and people's dislocation in the opposite direction in the afternoon. Pendular behavior is a relevant phenomenon that takes place in public transport networks because it may reduce the overall efficiency of the system as a result of the asymmetric utilization of the system in different directions. We apply this methodology to the bus transport system of Bras\'{i}lia, which is a city that has commercial and residential activities in distinct boroughs. We show that this methodology can be used to characterize the pendular behavior of this system, identifying the most critical nodes and times of the day when this system is in more severe demanded.\\

Published as: PHYSICAL REVIEW E 96, 012309 (2017)

DOI: 10.1103/PhysRevE.96.012309

\end{abstract}

\maketitle


\section{Introduction}

Recent research has revealed that human mobility patterns are not well described by random walk models~\cite{gonzalez2008}. Precise knowledge of these patterns is a fundamental step for promoting solutions that address the needs of real world issues such as traffic congestion~\cite{helb2001} and disease transmission~\cite{dal2013}.

As large databases are available and can be used to model human mobility, knowing how to differentiate between relevant and irrelevant pieces of information is important. Usually, two different approaches are considered~\cite{farland2016}, namely the data-driven approach and the hypothesis testing approach. The data-driven approach, which is common in science or machine learning literature, uses raw data or a transformed version of the data to determine distributions or to build models that can reveal interesting patterns. By contrast, the hypothesis testing approach, which is commonly  applied in economic theory or social science, uses the actual data to test a previously built model. Most approaches have considered the first paradigm~\cite{rhee2011,frank2013,haw2014}, and the gravity model~\cite{erlste90} that was recently reviewed in~\cite{Simini2012,farland2014} is a good example of the second approach. 

Our paper focuses on human mobility by means of public transport networks (PTN)~\cite{Latora2001,Latora2002,Sen2003,Sienkiewicz2005,Kurant2006,Xu2007,Chen2007a,Chen2007,Zhu2008,VonFerber2009,VonFerber2009,Caju2009,Caju2010} with the use of  a data-driven approach. While considerable literature deals with PTN, the literature that investigates topological representations of PTN is of fundamental interest here~\cite{Sen2003,Sienkiewicz2005}. In our paper, we use data from the bus transport system of Bras\'ilia, the Brazilian capital, which belongs to one of the federated units of the country, called Distrito Federal (DF). In this context, a relevant step is to mention a previous study~\cite{Chen2007} that analyzes the bus transport system of the four largest cities in China.

\begin{figure}
\centering
\subfigure{
\includegraphics[width=26em]{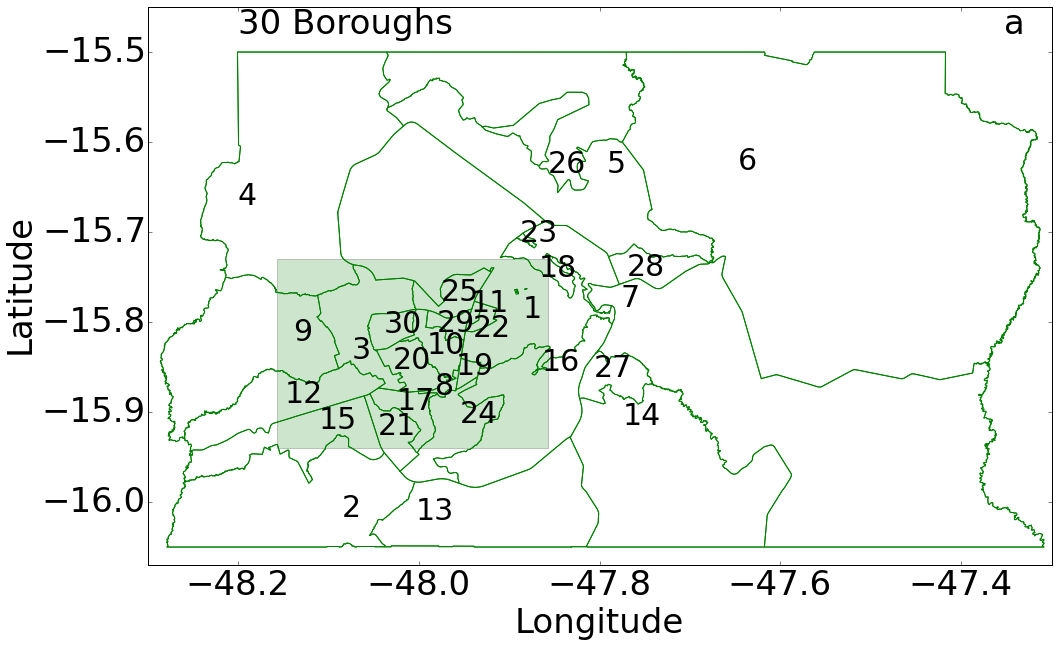}
\label{mapBoroughs}
}
\subfigure{
\includegraphics[width=26em]{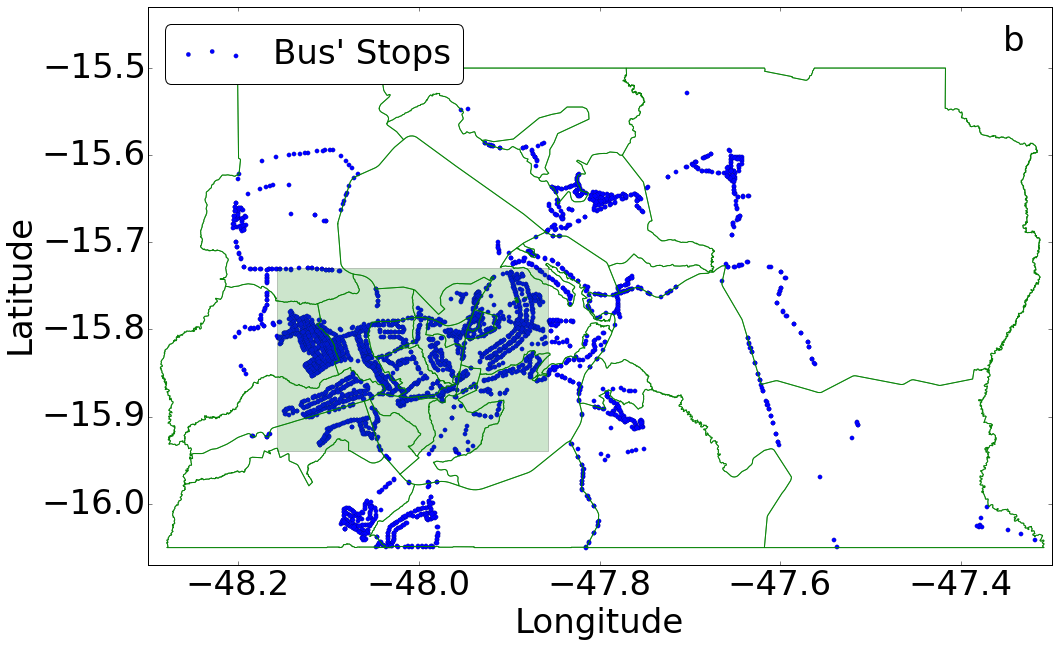}
\label{mapStops}
}
\caption{\subref{mapBoroughs} Administrative boroughs of the DF, numbered from 1
to 30 and \subref{mapStops} the 3673 bus stops of the public bus transport system. Number 1 refers to Bras\'ilia, which houses the federal government and is the center of the metropolitan area. Most of the population live in other boroughs, mainly in the southwest of Brasilia.
The rectangle encompasses most of the residential and job positions, and is the area analyzed in Fig.~\ref{pendularidadeDF}.}
\label{rasstopsDF}
\end{figure}

\begin{figure}
\centering
\includegraphics[width=26em]{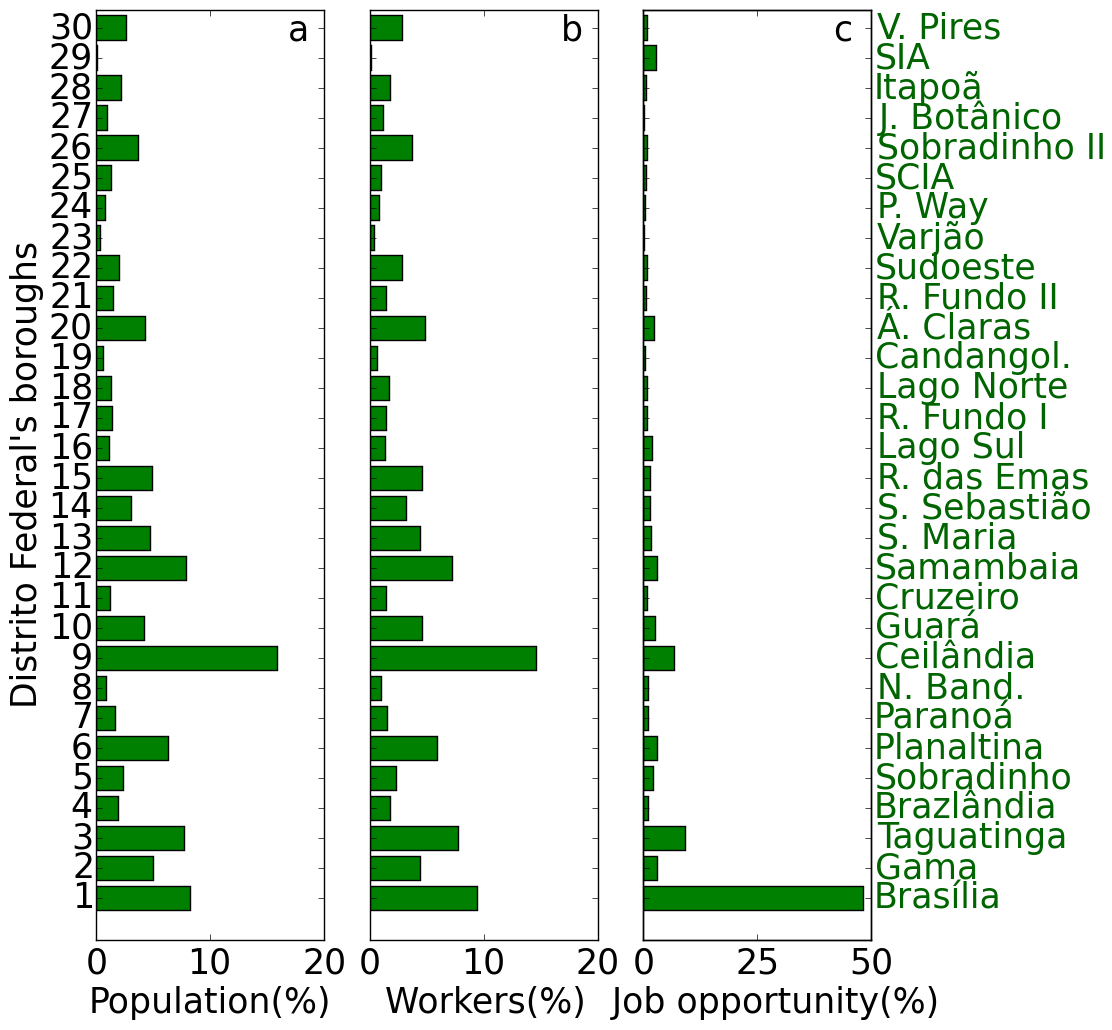}
\caption{Population distribution among the  DF's  boroughs. The geographical position of the boroughs are shown in Fig.~\ref{mapBoroughs}. The job positions are concentrated in Brasilia, and the commuting working is concentrated population in Ceilandia, Samambaia, and Taguatinga, which are found on the lower left side of the maps in Fig.~\ref{rasstopsDF} and~~\ref{pendularidadeDF}.}
\label{dataBoroughs}
\end{figure}

The DF comprises 30 boroughs [Fig.~\ref{mapBoroughs}], with Bras\'ilia being the one that concentrates public jobs (Fig.~\ref{dataBoroughs}). In 2011, census data (PDAD/DF-2011) estimate that the DF has a population of 2\,556\,149, with 1\,078\,261 employed individuals~\cite{Pad2012}.

Similar to many metropolitan areas, the DF has a heterogeneous population distribution with commercial and residential activities concentrated in distinct and far apart boroughs. Business and administrative areas  are highly concentrated in Bras\'ilia [borough 1 in Fig.~\ref{mapBoroughs} and \ref{dataBoroughs}], which has 477\,125 job positions, but with only 87\,736 of those occupied by local residents. The three geographically close boroughs of Ceil\^andia, Taguatinga, and Samanbaia [boroughs 3, 9, and 12 of Figs.~\ref{mapBoroughs} and ~\ref{dataBoroughs}] are the residences of most workers. This concentration of residences and jobs is reflected in the bus stop distribution, as shown in Fig.~\ref{mapStops}.

The existence of distinct residential and business neighborhoods results in a pendular movement of people between regions, with opposite travel directions in the morning and in the afternoon. A greater pendular disparity corresponds to a larger usage imbalance of the two routes' directions, thereby resulting in underutilization of some resources and reduced system efficiency. In this paper, we propose pendular movement measures that can help identify critical operation times. 

Our approach helps managers and users of the system identify the most vulnerable times of system operation, predict the best time to travel to a particular region, and improve the system. In particular, we found that the pendular behavior of the  bus transport system in Bras\'{i}lia can be assessed with only two measures that are used to summarize the properties of the center of mass (CM) of the distribution of the bus trips.

This  approach can also be used to understand the critical behavior of other networks, such as power grid networks~\cite{albertGrid2004,pagani2013} and financial loan networks~\cite{may2008,gai2011,haldane2011}, that are influenced by human actions that present bursts or other periodical temporal patterns. While power grid networks are subjected to daily human patterns, financial loan networks are subjected to monthly, yearly, and other temporal cyclical patterns of economic systems. Thus, our model can be helpful given the real-world importance of these networks and the fact that these networks are vulnerable to failures at all scales~\cite{helbing2013}.

\begin{figure}
\centering
\subfigure[$\,$ Transport system]{
\includegraphics[width=9em]{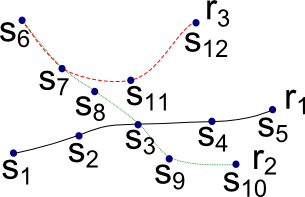}
\label{redeReal}
}
\subfigure[$\,$ L-space]{
\includegraphics[width=8em]{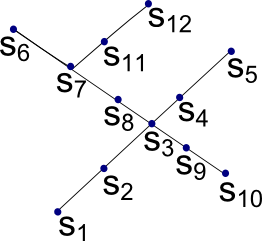}
\label{ptnRailBusL}
}
\subfigure[$\,$ C-space]{
\includegraphics[width=4.5em]{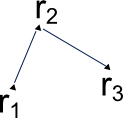}
\label{ptnRailBusC}
}
\subfigure[$\,$ P-space]{
\includegraphics[width=10em]{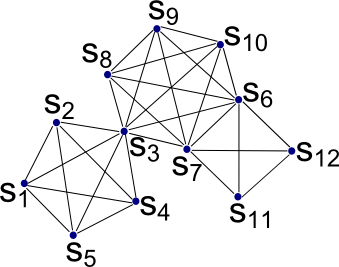}
\label{ptnRailBusP}
}
\subfigure[$\,$ B-space]{
\includegraphics[width=13em]{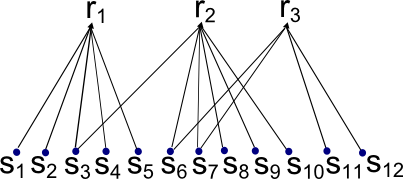}
\label{ptnRailBusB}
}
\caption{A public transport system is composed of routes that visit the same sequence of stops during each trip. The hypothetical system shown in ($a$) is composed of three routes, $r_1\dots r_3$, and twelve stops, $s_1\dots s_{12}$. Section \ref{sec:PTNspaces} discuss the four spaces that can be used when translating the system in a network. Networks ($b$)-($e$) are the representation of those spaces of the transport system shown in ($a$).}
\label{ptnRailBus}
\end{figure}

\section{PTN topology} \label{sec:PTNspaces}

Several different networks can be built from the same public transport system. Systems are composed of routes, trips, and stops. The stops may be georeferenced, and the trips usually define the instant when each stop is reached. It can also contain information, such as the number of passengers inboard. Meanwhile, the network is defined by its nodes and links, which, among other possibilities, can be directed or weighted.

The network topology will depend on which objects of the system are associated with the network nodes and on the rules that establish the links between nodes.
Four topological representations are commonly used to build a complex network out of
real public transport data \cite{Sen2003,Sienkiewicz2005,Chen2007,Chen2007a,VonFerber2009}.
\begin{description}
\item[P-space] Stops and stations are the nodes and an edge exists between two nodes if the nodes are simultaneously attended by at least one route~\cite{Sen2003,Sienkiewicz2005,Kurant2006,Chen2007}. 
\item[L-space] Graphs are similar to P-space, but each node is connected only to its immediate predecessor and successor on the routes~\cite{Sienkiewicz2005,Kurant2006,Chen2007}. 
\item[C-space] Routes are the nodes and an edge exists between two routes if they share at least one stop~\cite{Chen2007,Zhu2008}. 
\item[B-space] Routes and stops are both nodes of the network, and edges link routes to all their attended stops; however, no link exists between two stops or two routes~\cite{Chen2007,Zhu2008}. 
\end{description}
Fig.~\ref{ptnRailBus} depicts these four PTN networks, which result from a minimal and imaginary transport system.

Data from several cities were analyzed using these representations either referring to the  bus system~\cite{Chen2007,Xu2007,Chen2007a,Zhu2008,Sienkiewicz2005}, rail system~\cite{Sienkiewicz2005}, or subway system~\cite{Latora2001,Latora2002,Caju2009,Caju2010}. Only topological aspects were analyzed in these studies with no weighted or multi-edge links. Furthermore, the graphs were undirected and self-edges were not allowed.

Trains and subway systems may differ from bus networks in a way that affects the PTN topology. In most transport systems, train and subway stations can be reached by vehicles running in two directions, but all buses that attend a given stop travel along the same direction. Therefore, bus stops naturally separate the flow in both directions, whereas the directions become mixed in the stations of train networks. 

Pendularity is often a directional phenomenon. Thus, it will only be detected in a train system if measured over the directed edges. For that reason, we will restrict ourselves to the measurement of the edges pendularity even though it could also be determined over the nodes of bus networks.



A directed network can only be built if proper information regarding the direction of flow is available. If this is not the case, then pendularity studies would require a higher Fourier component to analyze the system, as discussed in the Appendix.


Another reason to focus on edges instead of the nodes is that the edges provide a more natural representation of the transit flow in the map. 

Among the representations discussed in this section, L-space is the most suitable for our analysis, because it naturally incorporates the trip's directionality between any two stops/stations. In addition, L-space can be immediately mapped to the street network where pendularity is observed, and it allows the construction of geographical maps that provide immediate visual information.

\section{Properties of the PTN elements}
\label{sec:properties}

\begin{figure}
\centering
\subfigure{
\includegraphics[width=25em]{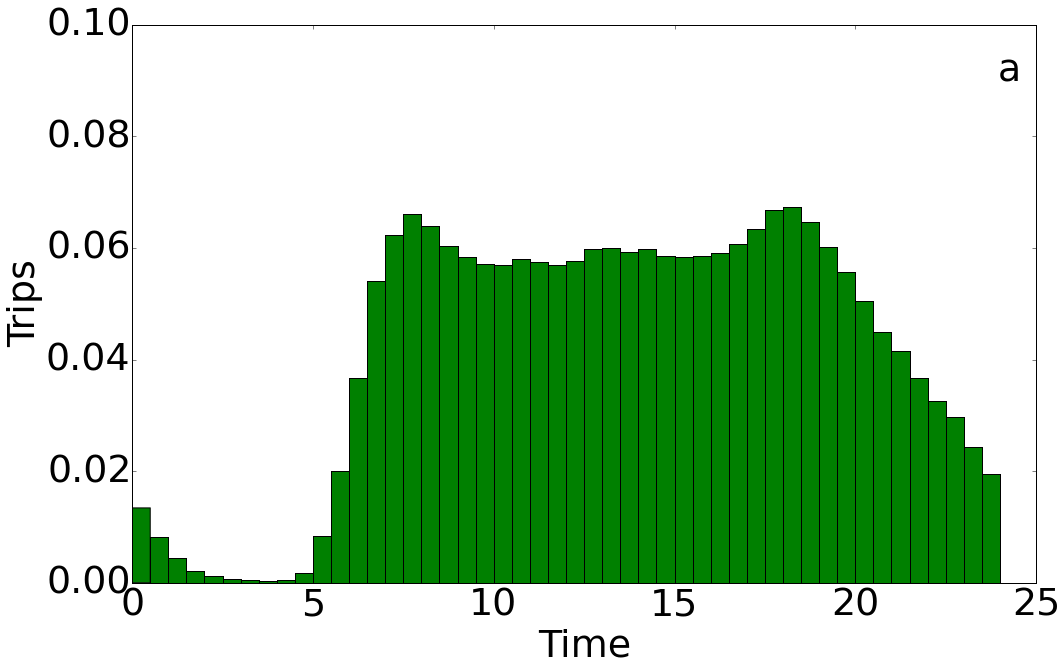}
\label{density1}
}
\subfigure{
\includegraphics[width=25em]{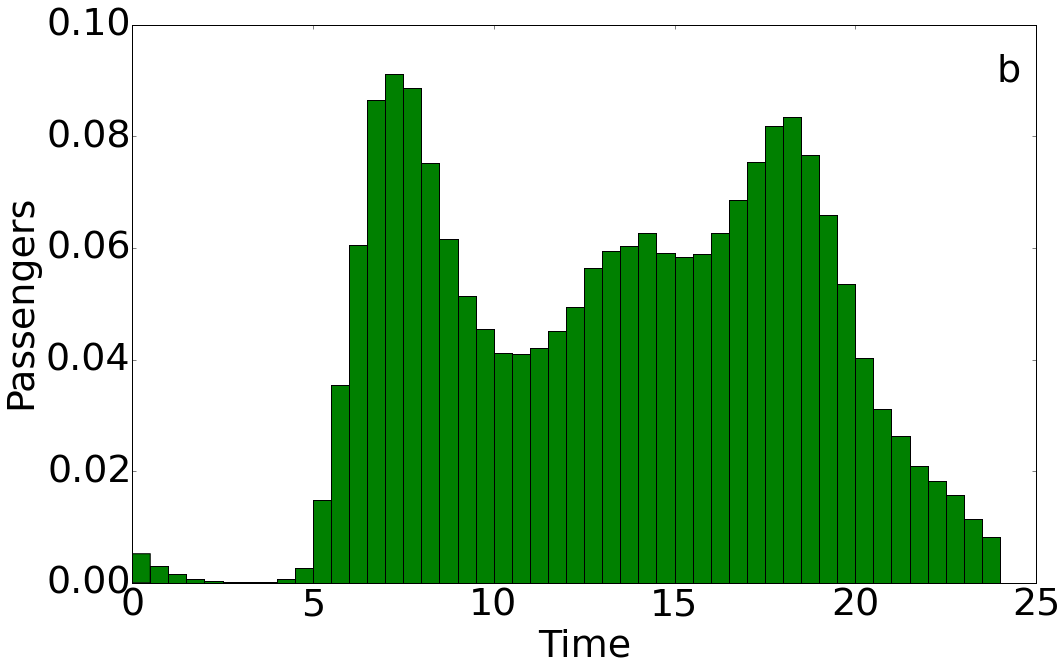}
\label{density2}
}
\caption{Normalized distribution along the day of every time a bus of the DF public transport system went through a stop on 13/Oct/2014, a typical weekday. In ($a$) the size of the bars is proportional to the number of buses stopping at each half-hour interval.  In ($b$) each stop is weighted by the number of passengers in the bus. The peaks related to the morning and the afternoon commuting are visible in both distribution although they are much more evident in the passengers' motion.}
\label{density}
\end{figure}

Most works on PTN deal with the network topology, which is built from the routes and stops of the public transport system. However further analysis of the PTN can be conducted if we consider other layers of information regarding, for example, the individual trips or the number of passengers. Although these pieces of information may look conceptually different, because the number of trips is related to the service supply and the number of passengers is related to the service demand, both pieces of information come from equilibrium data that can be represented as a kind of simultaneous equation model \cite{Wooldridge2001}. Therefore, individual trips roughly smooth the number of passengers. Although supply and demand peak during rush hours, as shown in Fig.~\ref{density}, demand has stronger peaks.

To better serve the users, the trips are not uniformly distributed along the day but concentrate during the hours with higher demand. It may appear that the number of trips per hour should be proportional to the average demand at that time of the day. However, this condition is neither feasible nor desirable.

A frequent property of public transport services in the DF is the highly directional demand during the rush hours, toward downtown in the morning and away from downtown in the evening. The vehicles that handle the morning demand must either stay parked downtown during the day or immediately return for a new trip. These returning vehicles may or may not be in service, and if they are not in service, then no trip is assigned to the empty vehicles. In any case, the demand is higher than the supply in the commute direction. Furthermore, even with very low demand, the vehicles' frequency may not be decreased below a given threshold to avoid excessive waiting times. 

Directionality is another piece of information that is usually not considered in the PTN analysis of stops. For routes and trips, the incoming flow of a given stop is, in general, equal to the outgoing flow and each of the flows is equal to half the total number of routes or trips of that stop. The only possible exceptions are the terminal stops. The situation is different with the directionality of passengers, because the number of passengers arriving at or leaving from a stop is not necessarily the same. However, the data available to us do not detail how the passenger numbers change within the trip; rather, the data show only the total number of passengers in each trip, and we assume that all passengers ride the whole trip. 



\section{Measures of pendularity}

\begin{figure}
\centering
    \subfigure[]{\includegraphics[width=22em]{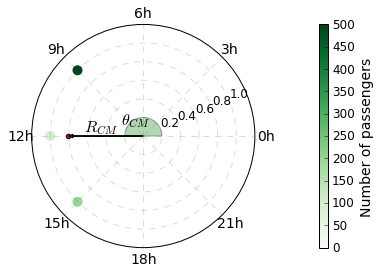}\label{fig:CMTrips}}
    \subfigure[]{\includegraphics[width=22em]{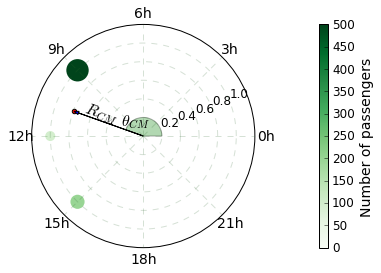}\label{fig:CMPass}}
\caption{A point along the circle of radius $r=1$ is assigned to each bus visiting a stop, the angle defined by the hour of the visit. Three stops are shown in the above figures, namely at 9 h, 12 h, and 15 h, from which the CM was calculated. In ($a$), every stopping of a bus has the same weight. In ($b$), the number of passengers is used as weight. The CM is attracted by heavier trips. The hour $h$ is connected to the angle $\theta$ by Eq.~(\ref{thetah}). }
\label{fig:CMcircle}
\end{figure}

Any measure that aims to analyze the distribution of a service around the day must reflect the 24-hour periodicity. To take into account the cyclical aspect of that problem, we associate an angle $\theta_i$ to the hour $h_i$ of each trip $i$ (24-hour clock), as defined by the equation
\begin{equation} \label{thetah}
    \theta_i = \frac{2\pi h_i}{24} \text{ rad}.
\end{equation}
The angle is used to place the trips along the circles of radius 1, as shown in Fig. \ref{fig:CMcircle}. 

If the events along the clock were placed in a straight line starting at 0 h and ending at 24 h, then instead of a circle, an artificial discontinuity exists, which would, for example, make a trip at 23 h 59 min be almost 24 h distant of a trip at 0 h 1 min. In this case, an equivalent approach is to explicitly consider the 24 h periodicity of the trip in a Fourier series, as discussed in the Appendix.

Once we define the position $\vec{r}_i$ of each trip along the aforementioned circle of radius 1, we can use the concept of CM to summarize some properties of that distribution.  The CM of the distribution of trips along the unitary circle, depicted in Fig.~\ref{fig:CMcircle},  is calculated as
\begin{align}
\label{rcm} 
\vec{R}_{CM}&=\frac{1}{M}\sum_{i=1}^N m_{i}\,\vec{r}_i, \\
\label{weight}
M &= \sum_{i=1}^N m_{i} .
\end{align}
The ``mass'' $m_i$ represents the trip importance. It may either be constant, which means that all trips are equally relevant, or it may be equal to the number of passengers on the bus, which means that the trip importance is proportional to the number of attended users.

Instead of using the vectorial notation, Eq. \ref{rcm} can be decomposed as
\begin{subequations}
\label{cmxy}
\begin{align}
X_{CM}&=\frac{1}{M}\sum_{i=1}^{N} m_{i}\, \cos\theta_{i}\ ,\\
Y_{CM}&=\frac{1}{M}\sum_{i=1}^{N} m_{i}\, \sin\theta_{i}\ .
\end{align}
\end{subequations}
From these components, we can write the radius and the hour of the
CM as
\begin{subequations}
\begin{equation}
R_{CM}=\sqrt{{X_{CM}}^{2} + {Y_{CM}}^{2}} \label{modRcm}\\
\end{equation}
\begin{multline}
h_{CM}=\frac{24}{2\pi} \arctan \frac{Y_{CM}}{X_{CM}} \\+ \label{hcm}
    \begin{cases}
         24 & \text{if } X_{CM}>0 \text{ and } Y_{CM}<0\\
        0 & \text{if } X_{CM}>0 \text{ and } Y_{CM}>0\\
        12 & \text{if } X_{CM}<0
    \end{cases}
\end{multline}
\end{subequations}
The hour of the CM is an average of the hours of the scheduled trips. Fig. \ref{fig:CMTrips} shows the CM of a symmetric distribution of trips. If the same trips are weighted by the number of passengers, then the hour of the CM moves toward the trip that more passengers attended [Fig. \ref{fig:CMPass}]. 

\begin{figure}
    \centering
    \subfigure[]{\includegraphics[width=18em]{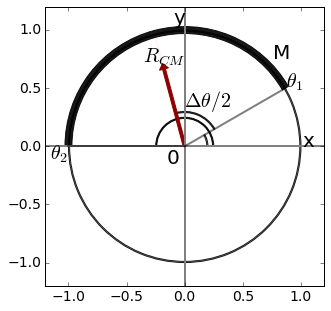} \label{fig:RCMdraw}}
    \subfigure[]{\includegraphics[width=18em]{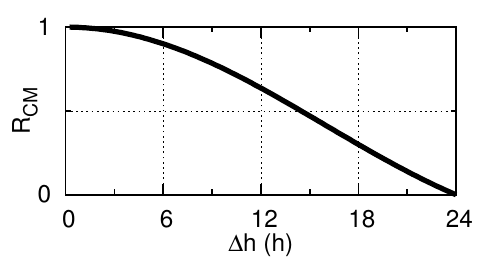}\label{fig:R_dh} }
    \caption{($a$) The CM of a continuous distribution is easily calculated and can be used to estimate the CM of a regular distribution of trips along the hours defining the angles $\theta_1$ and $\theta_2$. The $\vec{R}_{CM}$ points towards the mean value of these angles, and its length ($b$) is given by Eq.~(\ref{R_dh}). Larger values of $R_{CM}$ indicate more concentrated distributions. }
\end{figure}

The radius of the CM expresses how concentrated the trips or passengers are around a given hour. Its maximum value, $R_{CM}=1$, occurs when the trips are concentrated in one instant, for example, when only one trip is made per day. With trip weighting, the minimum value, $R_{CM}=0$, occurs when the trips are symmetrically distributed along the day, for example, if $N$ trips are made per day and they are separated by $24/N$ hours. If the $N$ trips are regularly distributed between the hours $h_1$ and $h_2$, with $h_1<h_2$, then the CM points to the mean value
\begin{equation}
    h_{CM} = \frac{h_1+h_2}{2}.
\end{equation}
An expression for $R_{CM}$ can be found if we use the continuous approximation ($N\rightarrow\infty$) shown in Fig.~\ref{fig:RCMdraw}. As the only goal of that distribution is to calculate $R_{CM}$, we symmetrize the distribution with respect to zero, thereby resulting in the integral
\begin{equation}
    R_{CM} = \frac{1}{\Delta\theta}\int_{-\Delta \theta/2}^{\Delta \theta/2}
    \cos(\theta) \, d\theta = \frac{2}{\Delta \theta}  \sin(\Delta \theta/2),
    \label{R_dh}
\end{equation}
where $\Delta\theta = 2\pi(h_2-h_1)/24$. Fig. \ref{fig:R_dh} shows how the radius $R_{CM}$ depends on the $\Delta h=(h_2-h_1)$. More concentrated trips imply smaller $\Delta h$ and bigger $R_{CM}$.

\section{Center of Mass of Public Transport Networks}

We start by drawing the PTN topology in the L-space. Next, we calculate and  attribute to each node (stop) or edge (route linking two stops) the following three properties: $M$, Eq. (\ref{weight}), the $h_{CM}$, Eq. (\ref{hcm}), and the $R_{CM}$, Eq. (\ref{modRcm}). 

With the use of data from the public transport system, the CM is calculated for each stop and its properties are assigned to the nodes.

Before calculating the CM of an edge in the L-space, the time of each trip that goes along the edge needs to be calculated. This time is defined as the mean time of the trip in the two nodes connected by the edge. After the calculation, the CM is obtained and the edge receive the properties $M$, $h_{CM}$, and $R_{CM}$.


When calculating these properties, the trips may all have the same weigh or the weigh may be proportional to the number of passengers, thereby resulting in two different network, which we will call {\em trips-weighted} and {\em passengers-weighted} networks. 

\begin{figure}
    \centering
    \includegraphics[width=26em]{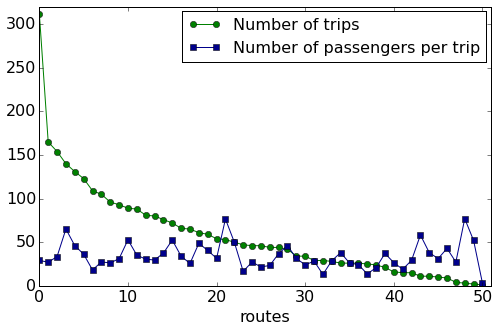} 
    \caption{Fifty routes of the DF public transport system with more trips in 13/Oct/2014, a typical weekday. The number of trips and the mean number of passengers per trip are shown for each route. The average and the standard deviation of the mean number of passengers per trip among these fifty routes are 34.25 and 14.85, respectively.  }
    \label{ViagensDia}
\end{figure}

In the following discussion, we will use data from the DF's transport system to build PTNs that incorporate the information from the CM analyis. The empirical data for this work were obtained from DFTRANS, the government agency for urban transport management. Data was collected for seven sequential days starting on 13/Oct/2014. No significant difference was found between the weekdays. Therefore we restrict our exposition to day 1. The day comprised 574 routes, 17\,479 trips, and 710\,765 passengers. Fig. \ref{ViagensDia} shows the 50 busiest routes of that day.

\begin{figure}
    \centering
    \includegraphics[width=26em]{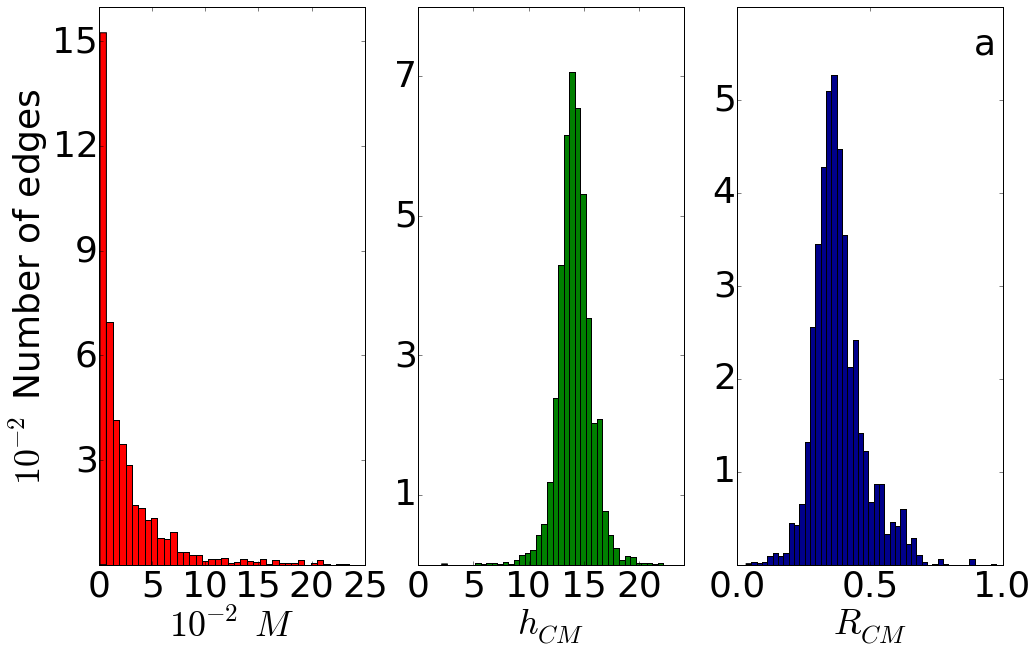}
    \includegraphics[width=26em]{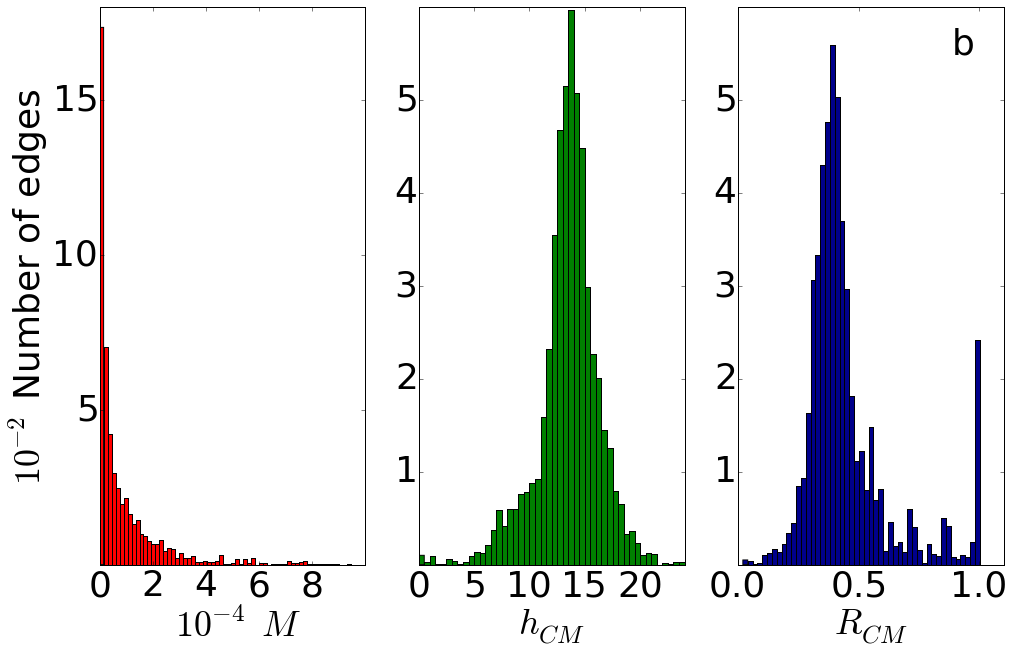}
    \caption{Distribution of $M$, Eq.~(\ref{weight}), $h_{CM}$, Eq.~(\ref{hcm}), and $R_{CM}$, Eq.~(\ref{modRcm}), of the network edges. (a) The upper plots are from the trips-weighted network and (b) the three lower plots are from the passengers-weighted network. The sizes of the bars are proportional to the number of \textbf{edges}. A significant number of edges with $R_{CM}=1$ exist in the passengers-weighted network, and are related to roads where most passengers commute in a short time interval even though no such concentration is observed in the trips-weighted network.}
    \label{fig:WRH_distrib}
\end{figure}

\begin{figure}
    \centering
    \includegraphics[width=26em]{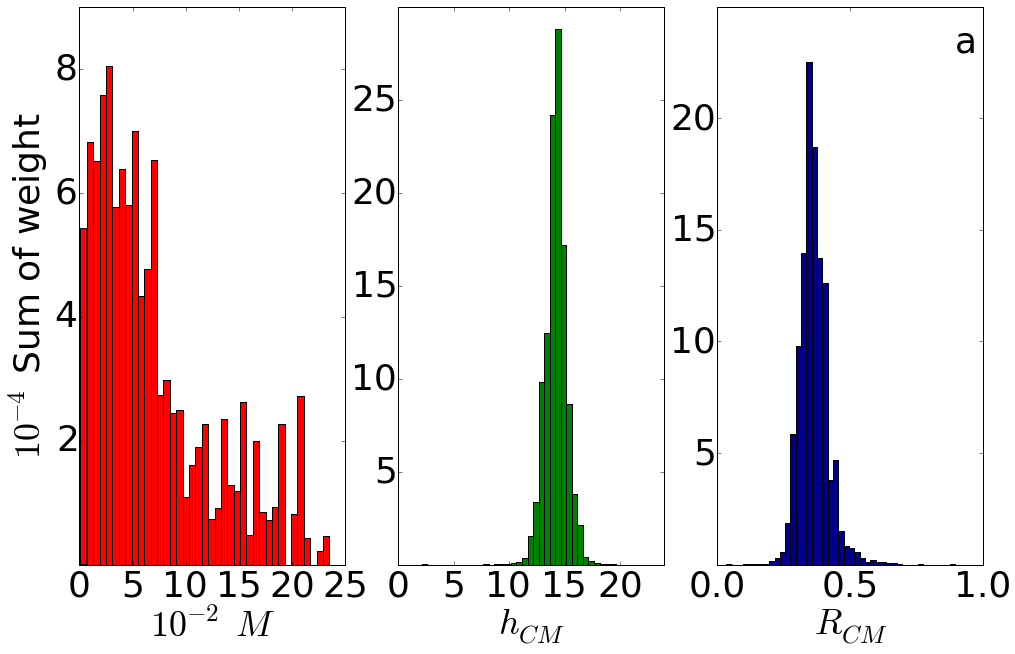}
    \includegraphics[width=26em]{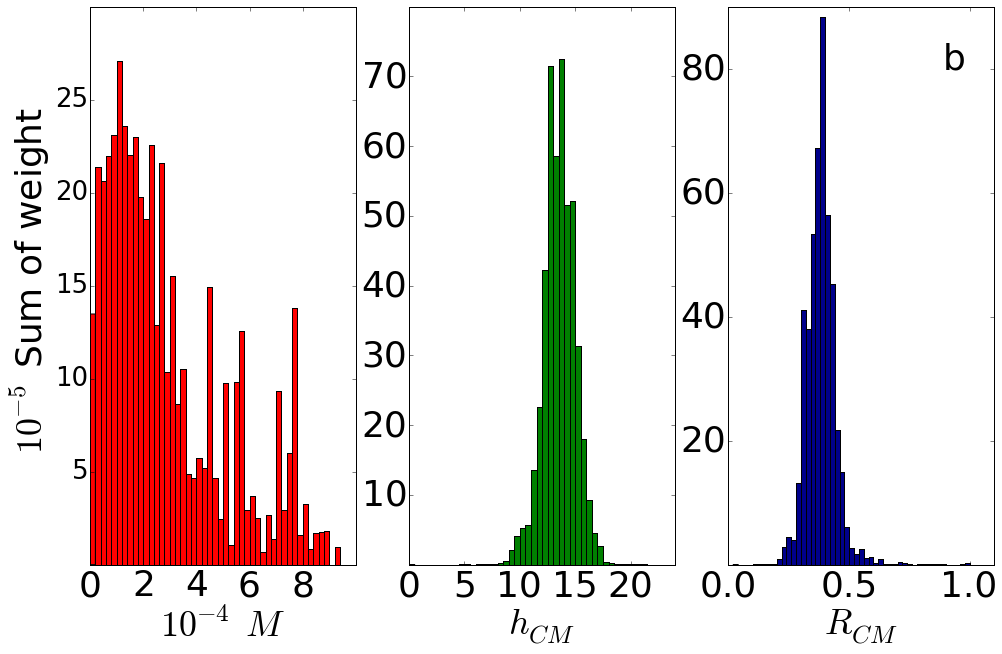}
    \caption{Distribution of $M$, Eq.~(\ref{weight}) , $h_{CM}$, Eq.~(\ref{hcm}), and $R_{CM}$, Eq.~(\ref{modRcm}), of the network edges. (a) The upper plots are from the trips-weighted network and (b) the three lower plots are from the passengers-weighted network. The sizes of the bars are proportional to the number of \textbf{trips} in the upper plots and to the number of \textbf{passengers} in the lower plots. A comparison with Fig.~\ref{fig:WRH_distrib} shows that the weighting can significantly affect the distribution, with the main effect being the reduced importance of edges with few trips or passengers. }
    \label{fig:WRH_Wdistrib}
    
\end{figure}

The distribution of $M$, $h_{CM}$, and $R_{CM}$ among the edges is shown in Fig.~\ref{fig:WRH_distrib}, where the height of each bar is proportional to the number of edges, i.e., all edges have the same weight. This edge weight must be clearly distinguished from the weight of the trips, which is used to compute the trips-weighted and the passengers-weighted networks.

Applying to the edges the same weighting schemes used in the trips makes sense. We perform this approach by defining the weight of the edge as equal to $M$, which is the sum of the weight of the trips on that edge (Eq.~\ref{weight}). This weighting was used to build the histograms of $M$, $h_{CM}$, and $R_{CM}$, as shown in  Fig.~\ref{fig:WRH_Wdistrib}. The weighting scheme is the only difference between Fig.~\ref{fig:WRH_distrib}, where all edges have the same weight, and Fig.~\ref{fig:WRH_Wdistrib}, where the weight of an edge is equal to the sum of weight of its trips.


Many irrelevant outliers present in Fig.~\ref{fig:WRH_distrib} disappear in the weighted histogram of  Fig.~\ref{fig:WRH_Wdistrib}. This evidence demonstrates the importance of including other properties, besides the topology, in the analysis of PTNs.

\begin{figure*}
    \centering
    \includegraphics[width=0.48\textwidth]{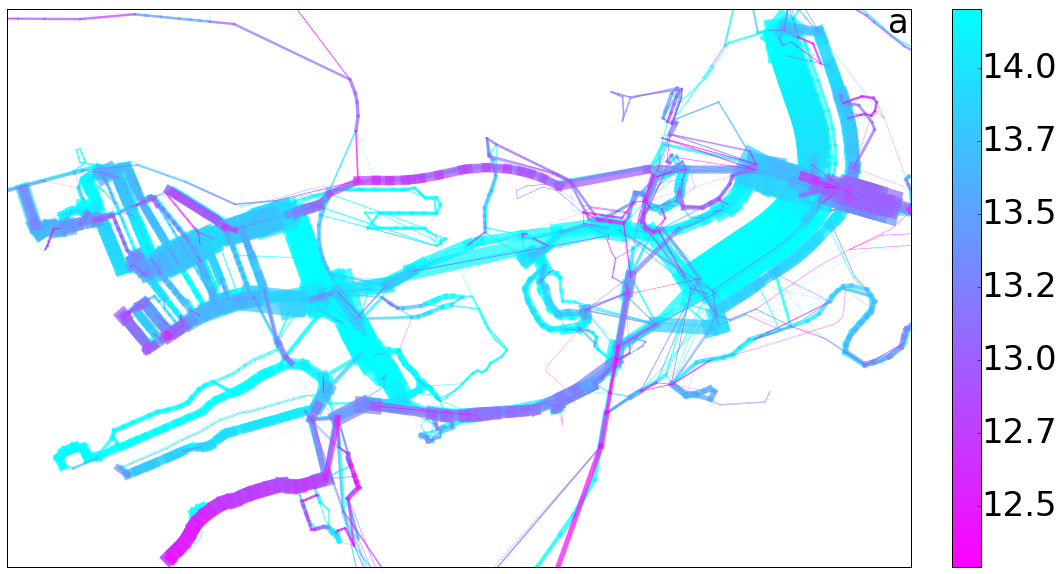}
    \includegraphics[width=0.48\textwidth]{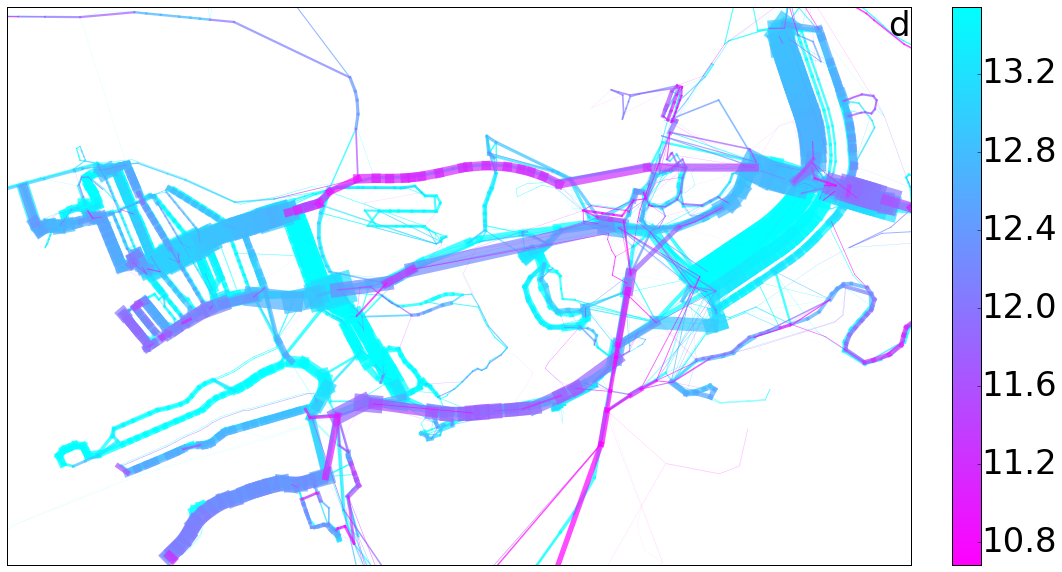}
    \includegraphics[width=0.48\textwidth]{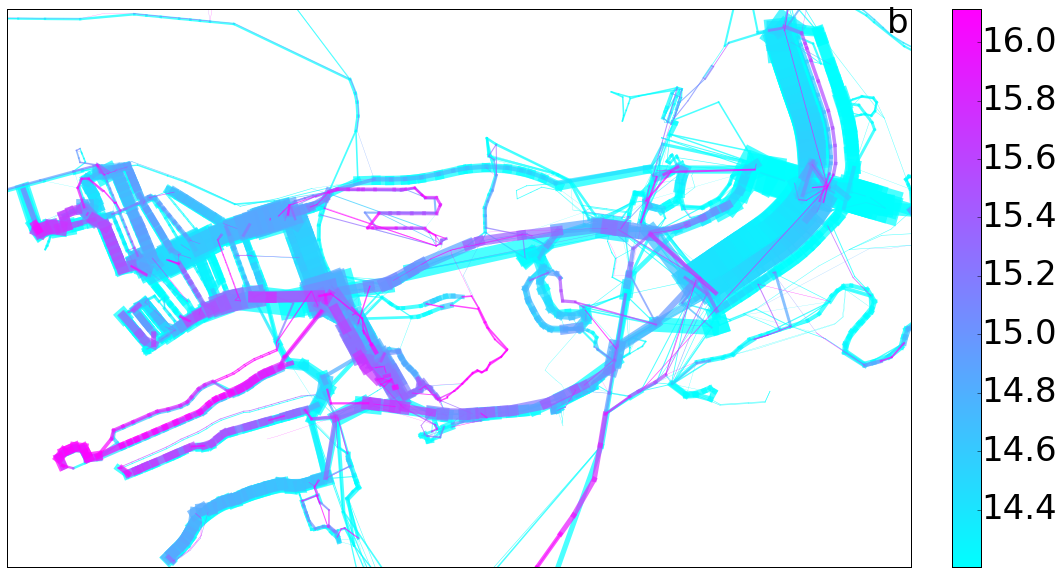}
    \includegraphics[width=0.48\textwidth]{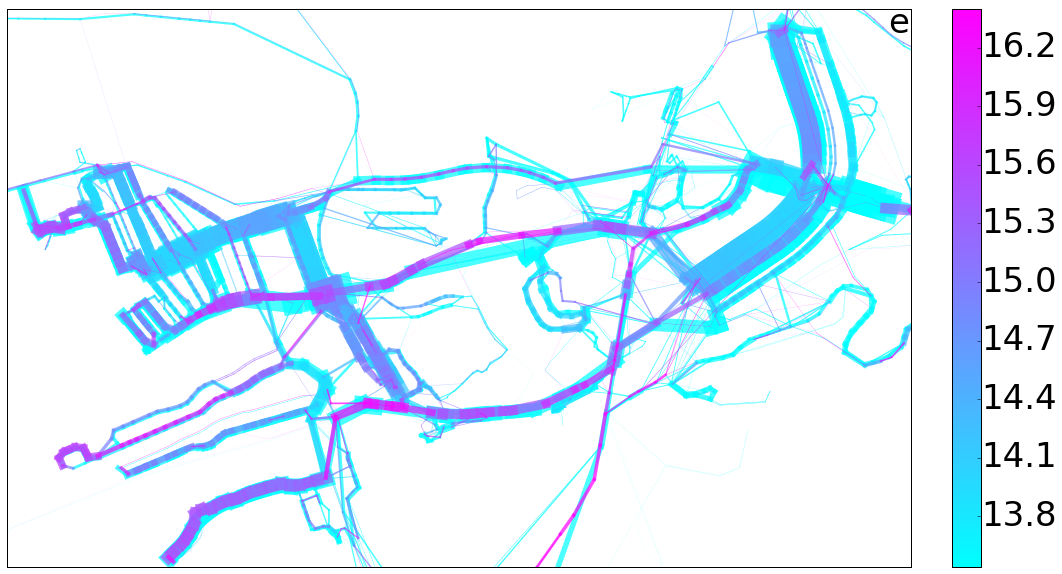}
    \includegraphics[width=0.48\textwidth]{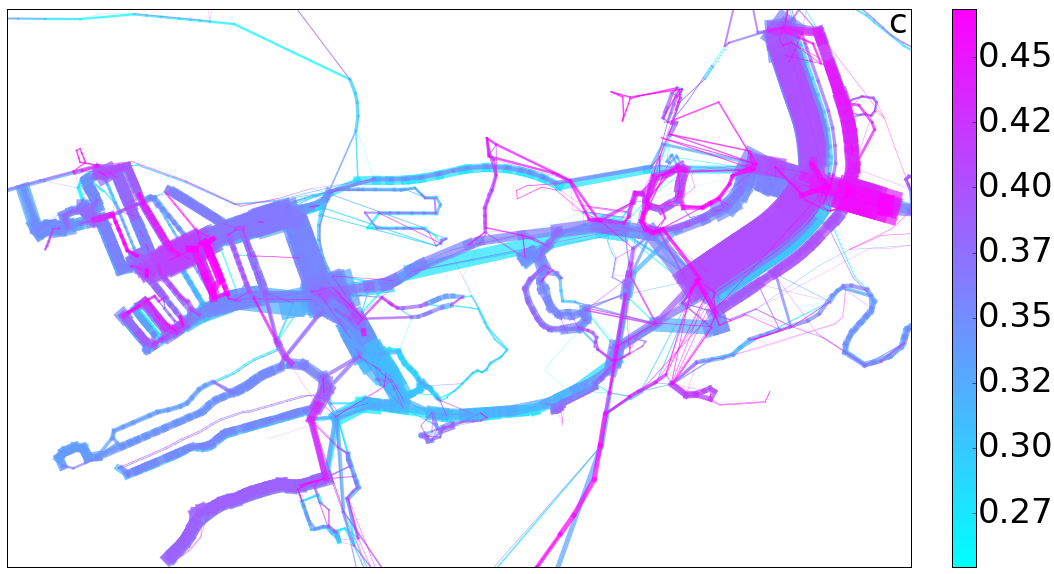}
    \includegraphics[width=0.48\textwidth]{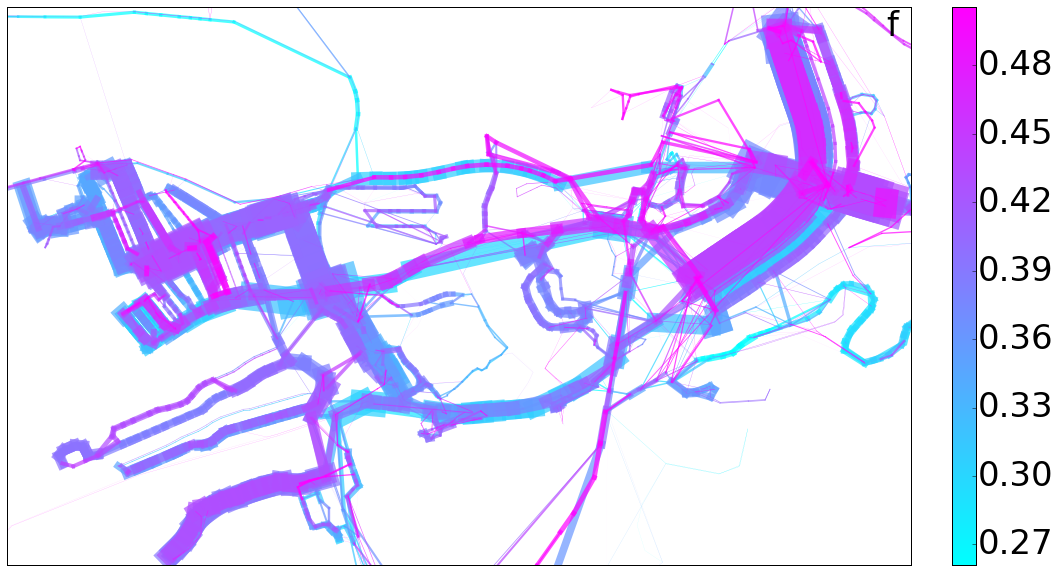}
    \caption{Pendularity of the most populated area of DF, highlighted in Fig.~\ref{rasstopsDF}. Pendularity is measured by the morning $h_{CM}$ (upper plots) and the afternoon $h_{CM}$ (middle plots), and the concentration of trips is measured by $R_{CM}$ (lower plots). The widths of the edges are proportional to the edge weight. The color of pendular edges is magenta and the the color of the non pendular edges is cyan. The three left plots use the number of trips as weight and the three right plots use the number of passengers as weight. (a) Morning $h_{CM}$ -- Trips; (b) Afternoon $h_{CM}$ -- Trips; (c) $R_{CM}$ -- Trips; (d) Morning $h_{CM}$ -- Passengers; (e) Afternoon $h_{CM}$ -- Passengers; (f) $R_{CM}$ -- Passengers.} \label{pendularidadeDF}
\end{figure*}

We classify the edges and the nodes as {\em morning} or {\em afternoon}  depending on whether their $h_{CM}$ value is greater or less than the average value. We plot these two sets of vertices in separate maps, shown in Fig.~\ref{pendularidadeDF}. 

From the distributions shown in Fig.~\ref{fig:WRH_Wdistrib}, we calculated the mean value ($\mu$) and the standard deviation ($\sigma$) of the $h_{CM}$ and $R_{CM}$. From these values, we defined the color interval of the maps of Fig.~\ref{pendularidadeDF}. For morning $h_{CM}$, we used the interval $[\mu_h-2\sigma_h,\mu_h]$. For the afternoon $h_{CM}$, we used the interval $[\mu_h,\mu_h+2\sigma_h]$. For the $R_{CM}$, we used the interval $[\mu_R-2\sigma_R,\mu_R+2\sigma_R]$.

These maps show each edge geographically located. The six maps are the combination of three measures, namely, morning $h_{CM}$, afternoon $h_{CM}$, and $R_{CM}$, and two weighting variables, namely, the number of trips and the number of passengers.

\section{Discussion}

Non-pendular edges are equally in demand for both directions during the morning and afternoon peaks, while pendular edges are more in demand for opposing directions in each peak. Therefore, $h_{CM}$ of non-pendular edges is closer to the average, and pendular ones have more extreme values. These extreme values are shown as the magenta edges of the $h_{CM}$ maps of Fig.~\ref{pendularidadeDF}.

As shown in Fig.~\ref{density}, the passenger distribution along the day fluctuates more widely than the corresponding trip distribution. The pendularity is related to distribution non-uniformity and is therefore expected to be more evident in the passengers-weighted network. However, this higher pendularity is not necessarily reflected in stronger colors in the passengers-weighted network of Fig.~\ref{pendularidadeDF}, because the color map is set to cover the range of values in a similar way.

Aside from distinct pendularity displayed in each network, distinct information is delivered by each map. The per-trip maps provide information regarding vehicle usage and traffic infrastructure load. The per-passenger maps reflect the attended population and the usage of the public transport installations.

The morning and afternoon traffic patterns are distinct. As can be seen in Fig.~\ref{density}, morning is traffic concentrated in the earliest morning hours and afternoon traffic is more spread out, because several activities are performed after regular working hours. These patterns explain the differences between the morning and afternoon maps.

The usefulness of $R_{CM}$ and $h_{CM}$ as measures of pendularity can be evaluated by using local traffic knowledge to analyze Fig.~\ref{pendularidadeDF}. The birdlike shape on the right side of the maps in Fig.~\ref{pendularidadeDF} is borough 1 (Bras\'ilia), where most jobs are located. On the left side, we can see the biggest residential area composed of boroughs 3, 9, 12, and 15 (borough information is presented in Figs.~\ref{rasstopsDF} and \ref{dataBoroughs}).

Three roughly parallel highways (Estrutural, EPTG, EPNB) connect these two areas. The upper highway is one-directional and reversible part of the day, whose counter-commuting traffic is diverted to the middle parkway. This counter-commuting flow reduces the pendularity of the middle parkway. The morning commute is mostly contained within the diverted traffic window, but the afternoon commute extends well after this window. Thus, the pendularity in the middle highway is higher in the afternoon. The strong pendularity of the lower and the upper highways is demonstrated by the magenta color of these routes in the $h_{CM}$ maps. 

Notably, the pendularity in the middle parkway is stronger in the passengers-weighted network than in the trips-weighted network. This finding indicates that a significant amount of people have a reason to take the middle parkway even if fewer trips are allocated there compared with the upper highway and will probably crowd the allocated vehicles.


A higher value of $R_{CM}$ in Fig.~\ref{pendularidadeDF}, corresponds to a more concentrated distribution of trips or passengers throughout the day. For example, the small magenta rectangle at the right side of these maps is the Esplanade of Ministries, an area exclusively occupied by federal government buildings. Most of the services there are provided from 8 h to 18 h, and the movement of people and buses concentrated in that interval explain the high value of $R_{CM}$ in that region.

We could further demonstrate the compatibility between local traffic knowledge and the pendularity expressed in the map. Such compatibility extends beyond the discussed cases. The presented discussion sufficiently states the reliability of $h_{CM}$ as a measure of pendularity. The evidence of pendularity is not as strong in the maps of $R_{CM}$, although it can contribute to the understanding of the daily traffic pattern.

\section{Conclusions}

In this paper, we introduced a methodology to identify and analyze networks that were subjected to pendular behavior. This methodology is able to identify the most critical nodes and times of the day when the behavior is critical. In particular, for the bus system of the DF, we showed that morning and afternoon traffic patterns were distinct for several edges.  The pendular behavior is proven by measuring the morning/afternoon asymmetry of the commuting movement. We rely on the establishment of a directional network that can distinguish the movement in both directions.  Pendular behavior results in a very low $h_{CM}$ on the edges pointing downtown because of the people's movement from home to work. Symmetrically, the value of $h_{CM}$ is high at edges pointing away from downtown.

We use $h_{CM}$ to classify the edges as morning or afternoon ones and compared the results with local knowledge of the city dynamics. Both $h_{CM}$ display the pendular behavior of the main commuting backbones. Furthermore, the separation allow us to evaluate the differences between the morning and the afternoon commutes.

The concentration of trips in the morning or in the afternoon should result in high values of $R_{CM}$ for pendular edges. Although a higher value of $R_{CM}$ should be correlated to higher pendular behavior, our color map of $R_{CM}$ does not highlight the known commuting backbones.

 This same methodology can be applied to identify the critical times of other types of networks such as financial networks or power networks. Future works may consider this line of research.

\appendix

\section*{Appendix: Fourier series of trip schedule}

Although Eq. (\ref{cmxy}) resulted from the definition of CM, it bears close resemblance to the first harmonic ($n=1$) of a Fourier series, whose coefficients are given by:
\begin{subequations}
\label{coef_fourier}
\begin{align}
x_n &= \frac{2}{T} \int_0^T f(x) \cos\frac{2\pi n t}{T}\, dt\\
y_n &= \frac{2}{T} \int_0^T f(x) \sin\frac{2\pi n t}{T}\, dt ,
\end{align}
\end{subequations}
with $T=24$~h.
That similarity becomes evident if one uses the Dirac's delta, $\delta(x)$, to write the trip scheduling as the distribution 
\begin{equation}
    \label{f_h}
    f(h) = \frac{1}{M} \sum_{i=1}^N m_i \delta(h- h_i).
\end{equation}
By substituting Eq. (\ref{f_h}) in Eq. (\ref{coef_fourier}) we obtain
\begin{subequations}
\label{harmonics}
\begin{align}
x_n &= \frac{1}{n\pi M} \sum_i m_i \cos\frac{2\pi n h_i}{24\text{ h}}\\
y_n &= \frac{1}{n\pi M} \sum_i m_i \sin\frac{2\pi n h_i}{24\text{ h}} .
\end{align}
\end{subequations}
From the above expressions we can define the amplitude of the harmonic $n$
\begin{equation}
    r_n = \sqrt{{x_n}^2+{y_n}^2} .
\end{equation}

The harmonic $n$ is related to trip distribution with a time interval of  $24/n$ h. For example, if two trips are made per day at 9 h and 17 h, i.e., separated by 8 h then, the trips would be related to $n=3$. The resulting coefficients are $r_n=1$ for $n$ as a multiple of 3 and $r_n=0.5$ for other values of $n$. The harmonic $n=3$ is the lowest of those with higher values, expressing the importance of the 24~h~/~3 = 8 h time interval of this trip schedule.

A more realistic situation would be to have 1 trip per hour from 7 h to 19 h, except from  8.5 h to 9.5 h, and from 16.5 h to 17.5 h, where the time interval between trips is 15 min. This situation would result in $r_1 = 0.55$, $r_3=0.47$, and smaller values for the other harmonics. The result not only reflects the trip concentration in half of the day but also the 8-hour time interval.

In most bus systems, vehicles travelling in opposite directions stop in distinct stops. This is not usually the case of trains systems. While unidirectional commuting stops peak either in the morning or in the afternoon, stops that attend to travels in both directions will present these two peaks. This double peak is probably better characterized by a Fourier harmonic with $n>1$.

A third peak may also commonly exist in the middle of the day, as can be seen in Fig. \ref{density2}. The presence of this peak is another situation in which a Fourier analysis with $n>1$ may be more suitable than $n=1$.

\end{document}